\documentclass[preprint,12pt,3p]{elsarticle}
\usepackage[english]{babel}
\usepackage{times}
\usepackage{hyphenat}
\usepackage{booktabs}
\usepackage{gensymb}
\usepackage{amsmath}
\usepackage{subfig}
\usepackage[figuresright]{rotating}
\usepackage{multirow}
\usepackage{setspace}
\usepackage{hyperref}
\usepackage{lipsum}
\usepackage{everysel}
\usepackage{lineno}
\usepackage[dvipsnames]{xcolor}
\usepackage{wrapfig}
\usepackage{adjustbox}
\biboptions{sort&compress}
\usepackage[cmintegrals,cmbraces]{newtxmath}
\usepackage[sfdefault,condensed]{roboto}
\usepackage[T1]{fontenc}

\begin{document}
	
\begin{frontmatter}
	
	\journal{arXiv}
	
	\title{\Large\textbf{Fabrication of bulk delta-phase Zirconium Hydride from Zircaloy-4 for use as moderators in microreactors}}
	
	\author[UCB,LANL]{D. Parkison}\corref{cor}\ead{dmsparkison@berkeley.edu}
	\author[LANL]{M.A. Tunes}\corref{cor}\ead{m.a.tunes@physics.org}
	\cortext[cor]{Corresponding authors, sharing first authorship.}	
	\author[SIGMA]{T.J. Nizolek}
	\author[LANL]{T.A. Saleh}
	\author[UCB]{P. Hosemann}
	\author[LANL]{C.A. Kohnert}

	\address[UCB]{Department of Nuclear Engineering, University of California at Berkeley, United States of America}
	\address[LANL]{Materials Science and Technology Division, Los Alamos National Laboratory, United States of America}
	\address[SIGMA]{Sigma Division, Los Alamos National Laboratory, United States of America}

\begin{abstract}
		\onehalfspacing
		\noindent The fabrication of bulk delta-phase Zirconium Hydride ($\delta$-ZrH) using Zircaloy-4 as a precursor is herein reported. Characterization using electron-microscopy methods indicate that the fabricated material is of a single-phase. Sn-rich segregation zones have been observed to form as a direct result of the hydriding process. These findings experimentally validate previous \textit{ab initio} calculations on the influence H incorporation in Zircaloy-4 constitutional elements such as Sn, Fe and Cr. The effect of hydriding and Sn segregation on pre-existing Zr(Fe,Cr)$_{2}$ Laves phases is also evaluated. Major implications on the development of moderators for use in microreactors within the nuclear industry are discussed. 
\end{abstract}


\end{frontmatter}

\textbf{Graphical Abstract}
	\begin{figure}[hb!]
	\centering
	\includegraphics[width=\textwidth,keepaspectratio]{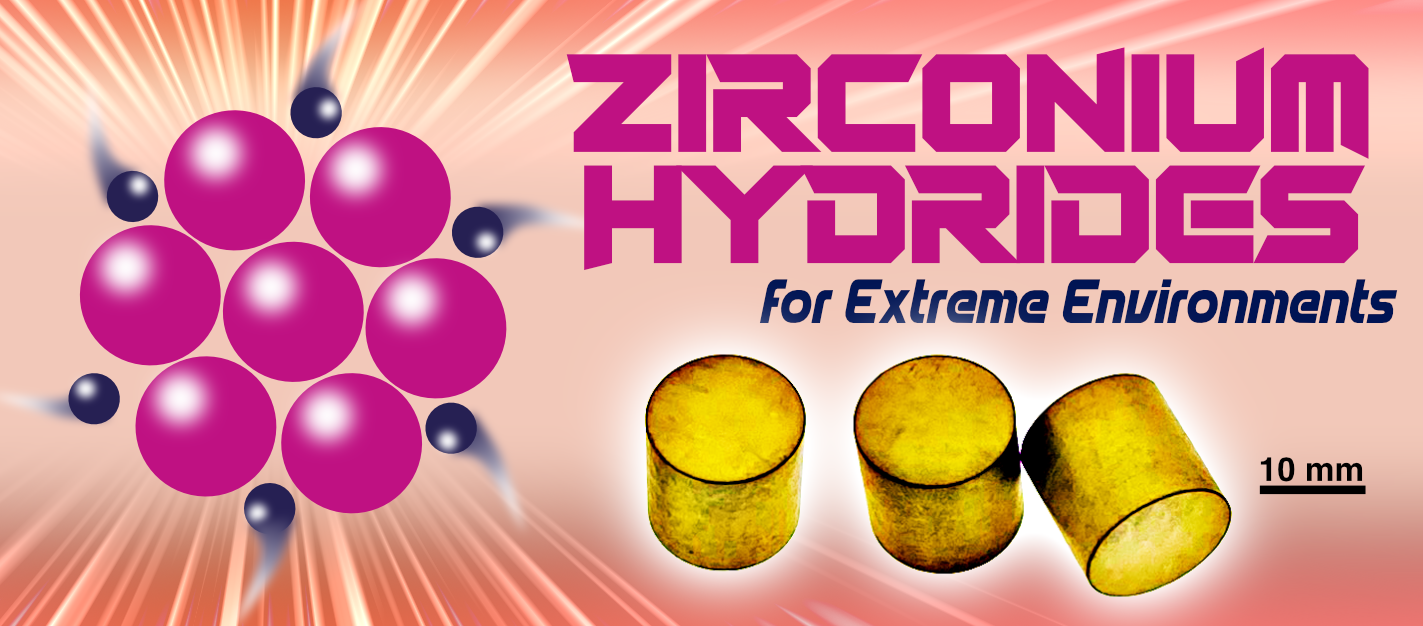}
	\caption*{}
	\label{GraphAbst}
\end{figure}

\newpage
\onehalfspacing

\noindent Transition metal hydrides are being studied as moderators for advanced nuclear concepts where reactor size is a critical factor \cite{mueller2013metal,wang2021advance}. Hydrogen -- with its mass being equal to that of a neutron -- is unique in its ability to moderate neutrons. Metal hydrides harness this ability whilst maintaining hydrogen atomic densities approximately equal to, or higher, than that of water over a wide range of temperatures  \cite{shivprasad2020advanced,shivprasad2020elastic,shivprasad2021thermophysical,wang2021advance}. Synthesis of metal hydrides for emerging technology applications is of paramount interest for the hydrogen, nuclear and materials communities \cite{cousins2023alloy,pandey2023numerical,atalmis2022effect,kazaz2022sorption,bhattacharyya2022multi,bannenberg2020metal}. In particular zirconium hydride (ZrH) is being pursued for its hydrogen content as well as its low thermal neutron capture cross-section \cite{mueller2013metal}. The synthesis of single-phase bulk transition metal hydrides remains an open challenge \cite{muta2018effect,wang2021advance}. In particular, literature on delta-phase zirconium hydride ($\delta$-ZrH) remains scarce when considering the hydride in a bulk form and unstudied when using a Zircaloy-4 precursor instead of pure Zr metal \cite{pinchuk1976nature,shcherbak1991radiation,kazarnikov1997effect,muta2018effect,oono2011irradiation,oono2013comparison}. 

The purpose of this study is to understand the effects of using Zircaloy-4 as a precursor in the fabrication of $\delta$-ZrH. Zircaloy-4 is an attractive precursor for bulk ZrH for a few reasons. First and foremost, as ZrH is being studied as a nuclear moderator, the neutron capture cross-section of the alloy precursor must be kept low. Hafnium has a very large thermal neutron capture cross section \cite{schemel1977astm}, and thus its concentration in the alloy must be controlled. Zircaloy-4 has a hafnium concentration that can be several orders of magnitude lower than other zirconium alloys \cite{schemel1977astm,pitrus1988determination,gordon1953spectrographic,horton1953spectrophotometric,ATI_Sheet}. The second benefit comes from the grain control of an engineering alloy. Zircaloy-4 has a smaller average grain size compared to other nuclear grade zirconium alloys -- namely ``Crystal Bar'', \text{i.e.} pure zirconium metal. The increase in the number of grain boundaries enhances diffusion \cite{balluffi1982grain} and it can therefore speed up the hydriding process.  Lastly, Zircaloy-4 is a well characterized material with  well established fabrication processes and commercial availability. This allows for large amounts of material to be purchased, with the consistent quality required for nuclear applications, from established vendors. 

For this work, nuclear-grade Zircaloy-4 was purchased from Allegheny Technologies Inc. (Albany, OR, USA) and machined into nominal 10x10 mm cylinders. The samples were cleaned with abrasive scouring pads and ethanol before hydriding. Samples were then hydrided according to known pressure-composition-temperature curves for the Zr--H system \cite{libowitz1962pressure}. All samples were hydrided in a single run using an Oxy-Gon model BC400 furnace (Oxy-gon Industries, USA)  with flowing ultra-high purity (UHP) H$_{2}$. The H/Zr ratio $\approx$ 1.66  was obtained by measuring the weight change after hydriding, assuming all weight gain was due to solid-state incorporation of H atoms into the Zr matrix via chemisorption \cite{ma2020cooperative}. Uncertainty in the H/Zr ratio was estimated to be around $\pm$0.02 and arises from systematic errors as well as the growth of thin oxide and nitride layers due to trace impurities in the UHP H$_2$ source-gas. A Bruker D2 Phaser tabletop x-ray diffractometer was used to confirm the phase purity of the hydrided samples.

Subsequently, the $\delta$-ZrH samples were then sectioned into small discs and polished using standard metallographic preparation ending in vibratory polishing using colloidal silica . After polishing, electron-transparent samples for transmission electron microscopy (TEM) were prepared with a Ga focused ion beam (FIB) system coupled into both Thermo Fisher Helios and Thermo Fisher Scios scanning electron microscopes (SEM).  The procedure for TEM sample preparation can be found elsewhere \cite{giannuzzi1999review}. Samples were subjected to electron backscatter diffraction (EBSD) in a Thermo Fisher Apreo SEM using an EDAX system. Electron-transparent samples were also analyzed using a Thermo Fisher Titan TEM 80-300 electron-microscope operating at 300 keV and featuring scanning transmission electron microscopy (STEM) capability and energy dispersive X-ray (EDX) spectroscopy suite.

\begin{figure}
	\centering
	\includegraphics[width=\textwidth,height=\textheight,keepaspectratio]{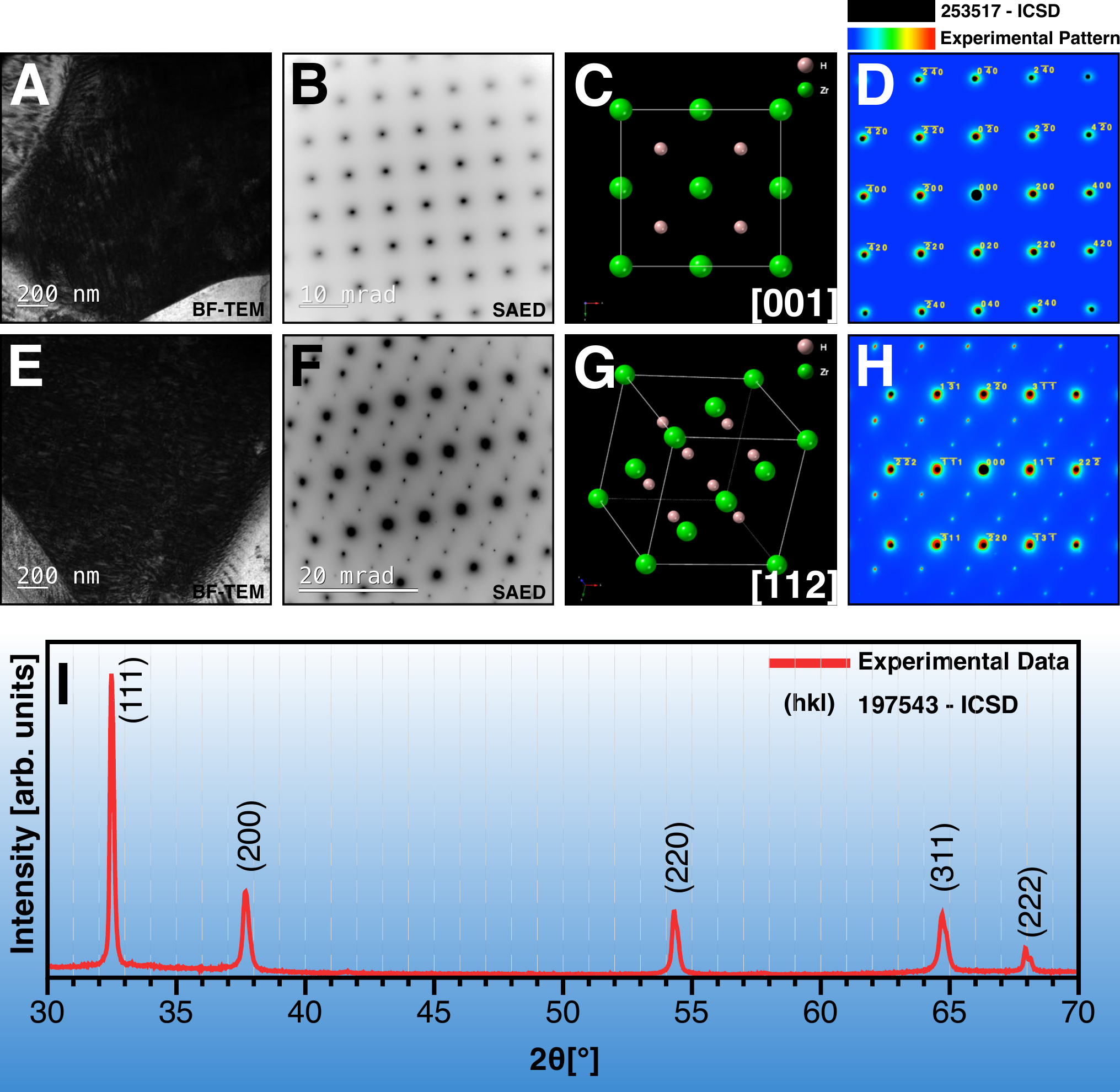}
	\caption{\textbf{Nano and microscale characterization of the bulk $\delta$-ZrH} | BF-TEM and SAED micrographs in A-B and E-F shows the bulk $\delta$-ZrH sample in two different zone axis. Simulation of crystal structures (C,G) and SAED patterns (D,H) in the CrystalMaker suite using data available in literature \cite{maimaitiyili2016observation,niedzwiedz199391zr} shows excellent agreement between experimental data and reference. The plot in I shows a XRD assessment of a bulk $\delta$-ZrH sample: the X-ray analysis indicates the bulk hydride is single-phase. Note: micrographs A and E were recorded in down-zone conditions giving a dark contrast appearance.}
	\label{fig01}
\end{figure}

After the hydriding process, TEM and XRD were used to assess whether the produced material was single-phase at both nano and macro scales, respectively. Figure \ref{fig01}(A-H) shows a crystallographic characterization performed within the TEM for $\delta$-ZrH samples in two different zone axis. The experimental SAED patterns in Figures \ref{fig01}(D) and \ref{fig01}(H) were overlayed with reference data available at the Inorganic Crystal Structure Database (ICSD) \cite{hellenbrandt2004inorganic} collection code 253517 \cite{maimaitiyili2016observation} and shows that the experimental data agrees with the reference data for single-phase $\delta$-ZrH. Figure \ref{fig01}I exhibits an experimental XRD pattern after hydriding where only $\delta$-ZrH diffraction peaks are observed. These diffraction peaks were indexed with ICSD crystallographic data collection code 197543 \cite{niedzwiedz199391zr}.

\begin{figure}
	\centering
	\includegraphics[scale=0.8,keepaspectratio]{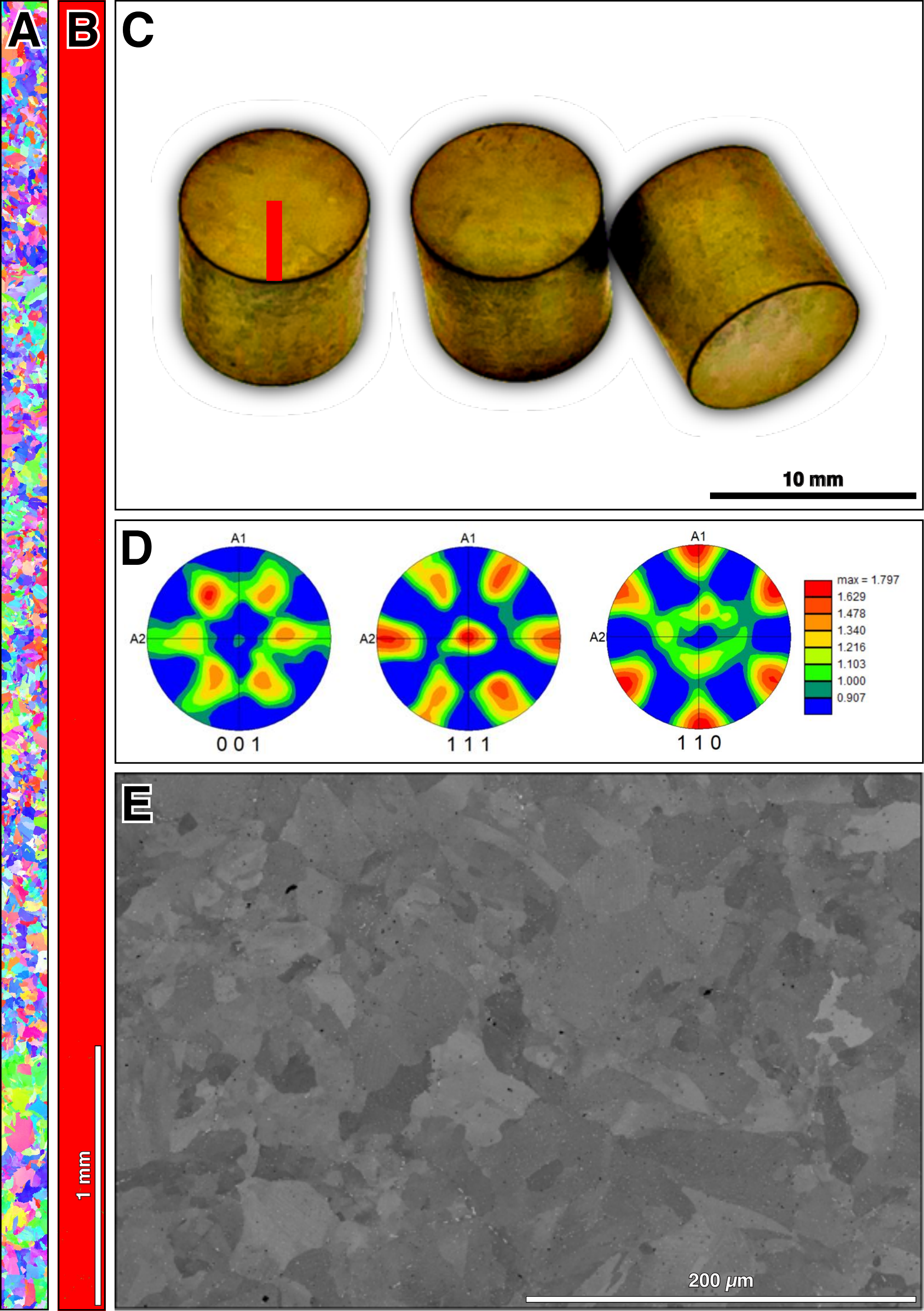}
	\caption{\textbf{The bulk $\delta$-ZrH} | EBSD confirms that the direct hydridring process fully transforms Zircaloy-4 into bulk $\delta$-ZrH. Figure A shows the inverse pole figure featuring approximately 4000 grains. The photograph in C shows the fabricated bulk pellet material and the small red rectangle indicates the EBSD region. The corresponding pole figures are shown in D. An overview of the $\delta$-ZrH microstructure as viewed by BSE is shown in E. Note: the scale bar in B also applies to A.}
	\label{fig02}
\end{figure}

Given that TEM and XRD suggest a single-phase bulk $\delta$-ZrH, EBSD was used to collect phase information on the samples after hydriding and also to assess cross-sectional phase purity. This technique is particularly useful in that it provides spatially resolved measures of phase fraction, grain structure, and crystallographic orientations for a large number of grains. Figure \ref{fig02}A shows an Inverse Pole Figure (IPF) map of a radial cross-section of a typical sample after hydriding. In this field of view, approximately 4000 grains were measured. The corresponding phase fraction map is shown in figure \ref{fig02}B. The approximate area covered by the EBSD analysis is shown by the red rectangle superimposed on a photograph of an as-fabricated sample in Figure \ref{fig02}C. Using EBSD it was found that the sample was >99.9\% $\delta$-ZrH with <0.1\% $\alpha$-Zr inclusions. While some of the points that indexed to HCP appear to be mis-indexed delta-phase with poor pattern quality, careful analysis of the Kikuchi bands revealed that multiple small inclusion do appear to be hexagonal in nature. The analysis of a hexagonal inclusion is shown in the supplemental material. Therefore, the samples are found to be over 99.9\% hydride. Average grain size was found to be approximately 12 $\mu$m (histogram in the supplemental material). Two large texture scans were conducted using a 10 $\mu$m step size and the results are shown in \ref{fig02}D. Analysis of this data reveals that the samples display a weak crystallographic texture (MRD<2) with clear orthotropic symmetry. It is worth noting that EBSD cannot distinguish $\delta$-ZrH from $\epsilon$-ZrH as a the c/a ratio change is not sufficiently large to appear in EBSD. However, EBSD and SEM backscatter imaging shows that the sample lacks the characteristic twinning of the pseudo-martensitic transformation indicative of $\epsilon$-ZrH as evidenced in \ref{fig02}E.

\begin{figure}
	\centering
	\includegraphics[width=\textwidth,height=\textheight,keepaspectratio]{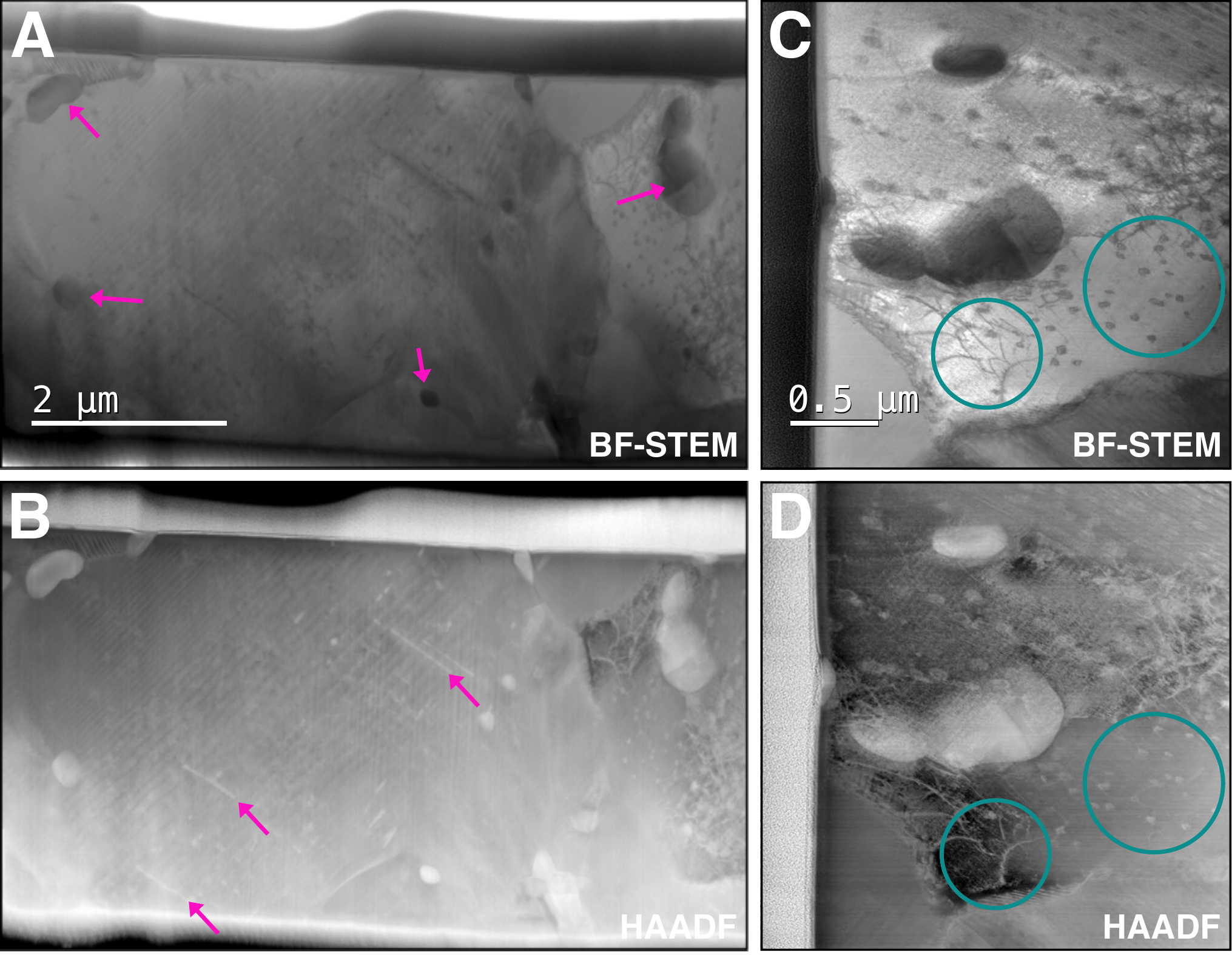}
	\caption{\textbf{STEM characterization of $\delta$-ZrH} | Micrographs A and C show the microstructure of $\delta$-ZrH under BF-STEM and micrographs B and D show the HAADF contrast counterpart. The characterization presents zones of SPPs observed after the hydriding process as indicated by arrows. Circles highlight the presence of defects such as dislocations.}
	\label{fig03}
\end{figure}

\begin{figure}
	\centering
	\includegraphics[width=\textwidth,height=\textheight,keepaspectratio]{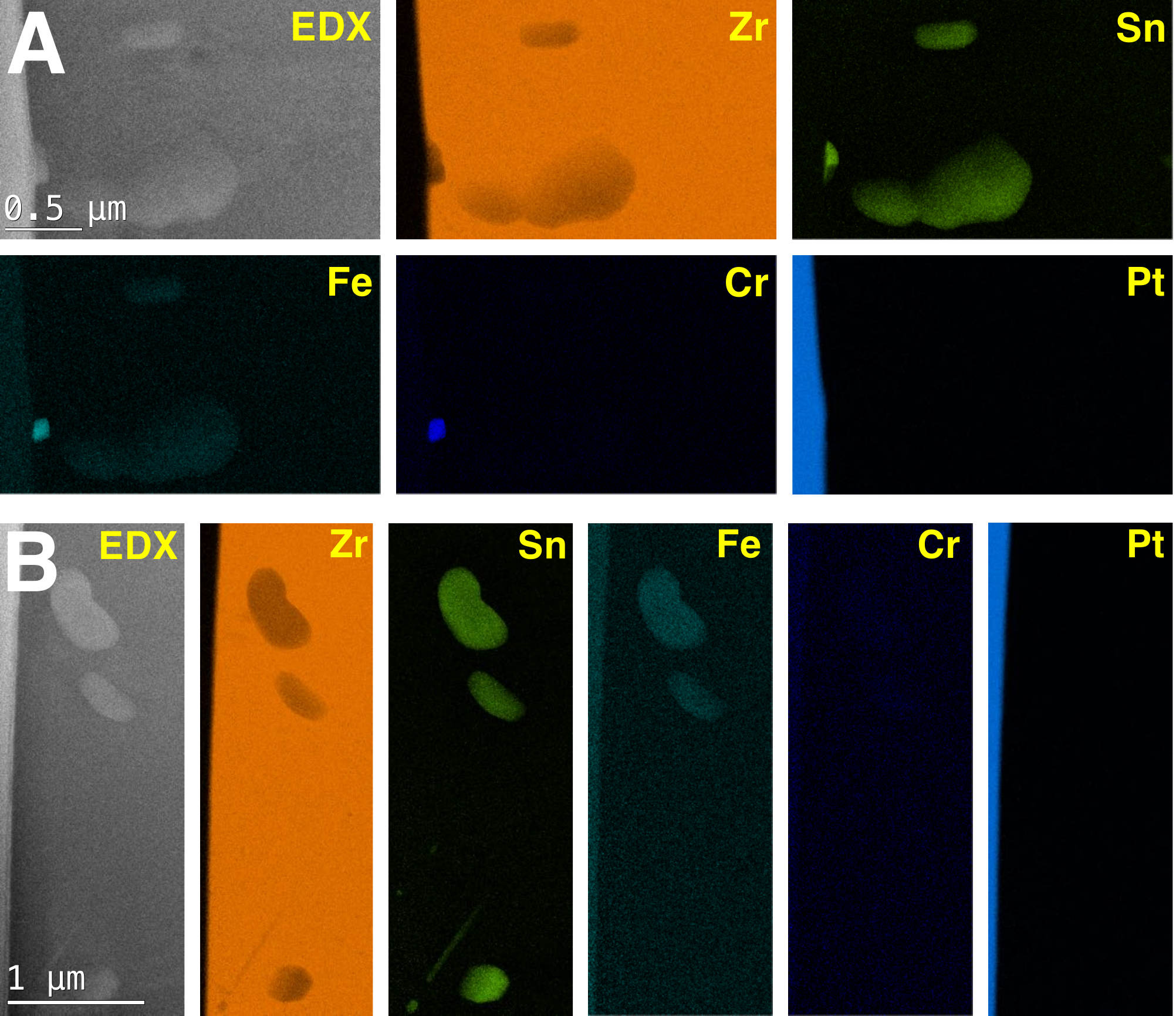}
	\caption{\textbf{Analytical STEM-EDX characterization of $\delta$-ZrH} | A series of two STEM-EDX maps taken at different magnifications and positions -- A and B -- is presented in the figure. Large SPPs ($>$400 nm) were observed to be rich in Sn, depleted in Zr, and with some minor additions of Fe and Cr. Sn was also observed to segregate in pre-existing Zr(Fe,Cr)$_{2}$ precipitates (Laves phase \cite{yang2013thermodynamic}).}
	\label{fig04}
\end{figure}

Microstructural characterization was performed with electron-microscopy methods such as TEM and STEM, as well as STEM-EDX. Figures \ref{fig03}A and \ref{fig03}B show a low-magnification bright-field STEM (BF-STEM) and high-angle annular dark-field (HAADF) micrographs, respectively, featuring multiple grains of an electron-transparent FIB lift-out lamella. Arrows in the micrographs point to zones of secondary phase particles (SPPs) present in the material after the hydriding process. Such SPPs were observed to have both rounded and needle-like shapes. Subsequent higher-magnification micrographs in Figures \ref{fig03}C and \ref{fig03}D show defects such as typical dislocations lines and loops within the microstructure of a hydride grain. Apart from SPPs and dislocations, no other defective structures were observed in the TEM lamellae. In order to determine the nature of such SPPs, STEM-EDX was performed and the results are shown in Figure \ref{fig04}(A,B). These SPPs were found to be Sn-rich with additions of Fe and Cr. As observed in the set of elemental maps in Figure \ref{fig04}A, pre-existing Fe,Cr-rich precipitates Zr(Fe,Cr)$_{2}$ -- known as the Laves phases  of Zircaloy-4 -- were also observed to be enriched in Sn.

It has been found that it is feasible to fabricate bulk $\delta$-ZrH from a Zircaloy-4 precursor. The predominately single-phase nature of these samples was confirmed through three different techniques spanning different length scales --- \textit{e.g.} SAED, EBSD, and XRD. Analysis using TEM/SAED across two different zone axis agreed well with the ICSD databases illustrating phase purity at the nanometer scale. Phase purity at the meso scale was determined through the use EBSD. By indexing the patterns against not just an FCC file ($\delta$-ZrH) but also against the hexagonal metal phase it was possible to see that the samples were >99.9\%  $\delta$-ZrH, with any detectable heterogeneity between the center and surface of a pellet.  EBSD analysis determined a very weak texture  (MRD<2)  with clear orthotropic symmetry. Lastly, XRD was performed to determine phase purity across the sample. The purity of these samples, is expressed in the lack of non $\delta$-ZrH peaks. Previously $\delta$-ZrH was fabricated in a bulk for by Muta \textit{et. al.} (2018) using pure Zr metal with a low 70 ppm Hf content \cite{muta2018effect}. Neutron diffraction analysis was used to prove that the sample was phase pure. EBSD was conducted in order to correlate hardness data as a function of crystal orientation. As the focus of the study was to correlate hardness data to crystallographic direction, fewer grains were analyzed and no claims about texture were made in the study.
  
In this study we primarily report that Sn-rich SPPs form in the bulk $\delta$-ZrH matrix after the hydriding process when fabricated from Zircaloy-4. As seen in the STEM-EDX maps in figure \ref{fig04}, it is worth emphasizing that Sn segregates out of the original Zircaloy-4 matrix (a Zr-based alloy made with $\approx$1.5Sn and minor additions of Fe and Cr \cite{de2015transmission,schemel1977astm}) and forms predominantly rounded and needle-like shaped Sn zones with minor addition of segregated Fe and Cr --- these latter belonging to pre-existing Zr(Fe,Cr)$_{2}$ Laves phases in Zircaloy-4 matrix.  This experimental observation on Sn-rich SPPs is particularly interesting considering the binary Zr--Sn phase diagram: at the concentration of $\approx$1.5Sn, where this element is fully dissolved into Zr, a binary alloy is formed in a single-phase structure that is HCP ($\alpha$-Zr) at low temperatures or BCC ($\beta$-Zr) at temperatures higher than 1136 K \cite{okamoto2010sn}. In both $\alpha$- and $\beta$-Zr, Sn is not expected to segregate out of solid solution solely as a result of thermal annealing \cite{sano1998atom}. The formation of Sn-rich SPPs appears to be a direct result of the hydriding process.

Recent efforts using computational science have been devoted to understanding the role of H atoms within transition metal lattices \cite{christensen2014effect,christensen2014h,christensen2015diffusion,varvenne2016hydrogen,maxwell2018molecular,wang2019investigating,wimmer2020hydrogen}. The driving force for the formation of Sn-rich SPPs during hydriding process can be understood using \textit{ab initio} predictions of Christensen \textit{et al.} \cite{christensen2014effect}. These authors calculated the energy necessary to insert a metal atom into the hydride lattice as a function of increasing H content for both Zircaloy-2 and -4. This quantity is defined as ``insertion energy''. Among all alloying elements studied, Sn showed the largest insertion energy varying from 200 to 520 kJ$\cdot$mol$^{-1}$ at H/Zr ratios from 0.5 to 2.0. At the $\delta$-ZrH stoichiometry, \textit{i.e.} H/Zr $\approx$ 1.6, the insertion energy for Sn metal into ZrH lattice is around of 300 kJ$\cdot$mol$^{-1}$. Based on our experimental findings, we conclude that rejection of the Sn atoms from $\delta$-ZrH and the formation of Sn-rich SPPs becomes more energetically favorable compared to remaining in solid solution.

In this present research, when nanoscale STEM-EDX elemental maps of the single-phase $\delta$-ZrH are performed, we noticed an apparent decrease of the pre-existing Fe,Cr-rich Laves phases in Zircaloy-4. This was a parallel observation from the results presented in figure \ref{fig04}. A STEM-EDX mapping of H-charged Zircaloy-4 is presented in the supplemental information file for comparison. This finding points to two hypotheses: (i) the hydriding process promote dissolution of such phases, or (ii) these Laves phases are transformed to Sn-rich segregation regions. For the known Zr(Fe,Cr)$_{2}$ Laves phase, the dissociation (or melting) temperature has been reported to be approximately 1773 K \cite{yang2013thermodynamic}. On the other hand, the maximum temperature for the synthesis of the $\delta$-ZrH is approximately 1173 K. This argument suggests that the first hypothesis is unlikely. The observation of Sn segregation into a single Laves phase precipitate in figure \ref{fig04}(A) suggests the second hypothesis deserves further investigation. When Sn is rejected from the Zr lattice due to hydriding, it has to enrich sinks within the microstructure. Trans-, intra- and inter-granular precipitates -- such as the Laves phase -- could act as sinks for Sn segregation. This observation indicates that further studies are needed to understand the thermodynamic effect of Sn into Laves phases in quaternary Fe--Cr--Zr--Sn systems.

In the context of H uptake in Zr cladding within nuclear reactors, the localized hydriding of $\alpha$-Zr also causes Sn segregation. It is shown in Mouton \textit{et al.} that Sn segregates at the hydride/metal interface as studied via Atom Probe Tomography (APT) \cite{mouton2021hydride}. Sn was found to be enriched at stacking faults found ahead of the hydride/metal interface. Udagawa \textit{et al.} showed that Sn lowers the stacking fault energy in the Zr metal and proposes that these planar defects facilitate the nucleation of the hydride  \cite{udagawa2011effect} and this has also been corroborated by Chakraborty \textit{et al.} \cite{chakraborty2022effect}. 

The effect of Sn segregation could have wide ranging impacts on the stability of ZrH, especially within the contexts of irradiation and corrosion resistance, however, reports on Sn-rich SPPs is limited to Zr-based alloys studies and not in bulk hydrides. In the latter case, it has been reported that Sn segregation accelerates nodular corrosion processes through interactions with the metal-oxide interface \cite{xie2017novel}. Additionally, Sn segregation has also been reported as a direct effect of irradiation of Zr cladding \cite{shen2014proton,harte2017effect,cockeram2019atom,yu2023stem}. Future studies should aim to evaluate the influence of Sn-rich SPPs in bulk ZrH as fabricated using Zircaloy-4 as a precursor.

We have reported on the feasibility to fabricate bulk, high-quality, >99.9\% phase pure $\delta$-ZrH using Zircaloy-4 as the precursor material. The advantages of using the alloy instead of Zr metal have been outlined. A significant finding, obtained through careful TEM analysis, is that Sn-rich precipitates are present within the delta-phase matrix of the fabricated hydride material. The nature of such Sn-rich SPPs formation and impact on the overall response of the bulk $\delta$-ZrH as a neutron moderator material is pending of further research. The results in this paper also experimentally validate previous \textit{ab initio} computational results presented by Christensen \textit{et al.} \cite{christensen2014effect} on influence of major and minor alloying elements into Zr and ZrH systems. In order to better evaluate the homogeneity, distribution and concentration of H atoms within hydrides, novel spectroscopy techniques within the TEM are currently under development \cite{tunes2019site,badr2023trigonal} which may also help the efforts to better characterize hydrides in the future. 

\section*{Acknowledgments}

\noindent Research presented in this article was supported by the Laboratory Directed Research and Development (LDRD) program of Los Alamos National Laboratory primarily under project number 20220597ECR. MAT acknowledges support from the LDRD program 20200689PRD2. This work was supported by the U.S. Department of Energy, Office of Nuclear Energy under DOE Idaho Operations Office Contract DE-AC07- 051D14517 as part of a Nuclear Science User Facilities experiment. This work was supported by the NASA Space Technology Mission Directorate (STMD) through the Space Nuclear Propulsion (SNP) project.

\section*{Disclosure statement}

\noindent The authors declare no conflict of interest.

\section*{CRediT author statement}
\begin{itemize}
	\item  MAT and DP - Conceptualization, Methodology, Formal Analysis, Investigation, Data Curation, Visualization, and Writing - Original Draft.
	
	\item CAK and TJN - Fabrication of the specimens, Methodology, Validation, Formal Analysis, Investigation, Writing - Review \& Editing.
	
	\item TAS - Funding Acquisition, Writing - Review \& Editing.

	\item PH - Supervision of DP, Validation, and Writing - Review \& Editing.
	
	\item CAK - Supervision of MAT and DP, Conceptualization, Project Administration and Funding Acquisition. 
	
	\item Note: DP and MAT share the first authorship of this paper. 
\end{itemize}

\section*{Data availability}

\noindent All the data is already presented in the manuscript and supplemental materials. Raw data is available upon request.

\bibliographystyle{elsarticle-num}
\bibliography{biblibrary}

\end{document}